\documentclass[twocolumn,prb,amsmath,superscriptaddress,nofootinbib,amssymb]{revtex4}
\usepackage{graphics}
\usepackage{amssymb}
\usepackage{graphicx}
\usepackage{times}
\usepackage{epsfig}          

\begin{document}

\title{Interplay of electronic structure and atomic ordering on surfaces:\\
    Momentum-resolved measurements of Cs atoms adsorbed on a Ag(111)
    substrate}

\author{Hendrik Bentmann}\affiliation{Experimentelle Physik VII and R\"ontgen Research Center for Complex Materials (RCCM), Universit\"at W\"urzburg Am Hubland, 97074 W\"urzburg, Germany}\affiliation{Karlsruhe Institute of Technology KIT, Gemeinschaftslabor f\"ur Nanoanalytik, D-76021 Karlsruhe, Germany}
\author{Arne Buchter\footnote{present address: Department of Physics, University of Basel, Klingelbergstr. 82, CH-4056 Basel, Switzerland}}\affiliation{Experimentelle Physik VII and R\"ontgen Research Center for Complex Materials (RCCM), Universit\"at W\"urzburg Am Hubland, 97074 W\"urzburg, Germany}
\author{Friedrich Reinert}\affiliation{Experimentelle Physik VII and R\"ontgen Research Center for Complex Materials (RCCM), Universit\"at W\"urzburg Am Hubland, 97074 W\"urzburg, Germany}
\affiliation{Karlsruhe Institute of Technology KIT, Gemeinschaftslabor f\"ur Nanoanalytik, D-76021 Karlsruhe, Germany}

\date{\today}
\begin{abstract}
Surface-state mediated interactions between adsorbates on surfaces can be exploited for the fabrication of self-organized nanostructures such as two-dimensional superlattices of adatoms. Using angle-resolved photoemission we provide experimental evidence that these interactions can be drastically modified by adsorbate-induced alterations in the surface potential barrier. This, in turn, will cause significant changes in the ordering of the adsorbates. For the studied case example of Cs adatoms on Ag(111) our momentum-resolved measurements reveal the surface state Fermi wave vector to be increased by as much as $\sim$100\% for coverages around 0.03~ML. Our results unravel the origin for the hitherto puzzling and unexpectedly small lattice constant in the adatom superlattice observed for this system.     
\end{abstract}
\maketitle
The self-assembly of atoms, molecules or clusters on surfaces is a prerequisite for the bottom-up approach to the fabrication of ordered low-dimensional nanostructures. The interest in such processes is widespread: it reaches from improvements in the growth quality of organic molecular thin films for enhanced device performances \cite{Cho:07} to the design of model systems with tunable interaction parameters for the investigation of new electronic or magnetic effects.\cite{Schaefer:11.10,Bode:07.5,gambardella:09.3,Sakamoto:09.3,Ortega:11.8} The structural ordering of adsorbates is often determined by a complex interplay of competing mutual interactions which may act directly between the adsorbates or via the supporting substrate.\cite{Berndt:09,Kumpf:09.2,otero:11.11} The controlled preparation of tailored structures thus requires understanding of the involved microscopic mechanisms.\cite{barth:05,forster_systematic_2006} Here, we address specific types of interactions acting between adatoms and molecules on surfaces which are mediated by surface state electrons.\cite{Weiss:12.01} The mediated interactions can be electronic, as found for the surfaces of different metal substrates \cite{repp_substrate_2000,knorr_long-range_2002}, but also magnetic, as predicted for impurities with magnetic moments on topological insulator and metal surfaces.\cite{Zhang:09.4,Stepanyuk:04.8} As we will show, such interactions are not necessarily a genuine property of the substrate surface but instead can exhibit substantial dependence on the adsorbate type and coverage.

Surface-state mediated interactions result from the scattering of surface electrons at adatoms which gives rise to spatial oscillations in the local density of states (LDOS) of the surface state around the scattering impurity.\cite{crommie_imaging_1993,Hyldgaard:00.1} This effect induces an oscillatory interaction potential between two given adatoms whereas the oscillation period is directly linked to the surface state Fermi wave vector of the particular substrate. Scanning tunneling microscopy (STM) experiments very succesfully confirmed these predictions for several adatom species on different noble metal substrates hosting Shockley-type surface states.\cite{repp_substrate_2000,knorr_long-range_2002} A particularly interesting consequence of the surface-state mediated interaction potential is the possibility to create well-ordered adatom superlattices, potentially useful for studying magnetic coupling in low dimensions (see for example Ref.~\onlinecite{Ternes:10.1} for a recent review). Experiments for Ce superlattices on Ag(111) and Cu(111) corroborated a clear correspondence between the equilibrium interatomic distances in these structures and the respective surface state Fermi wave vector.\cite{silly_creation_2004,Schneider:09.6} On the other hand, Cs superlattices on the same substrates were revealed to show markedly reduced interatomic distances compared to the respective Ce counterpart \cite{Berndt:06.7,ziegler_scanning_2008}, indicative of additional, unexplored effects taking place in this case.

In this contribution we demonstrate that the surface-state mediated potential on a particular substrate can be altered significantly by the presence of adatoms and moreover can be tuned as a function of the adatom coverage. This effect arises from adsorbate-induced changes in the surface potential barrier which give rise to drastic modifications in the Fermi wave vector and the binding energy of the surface state electrons. To access these important quantities experimentally we employed angle-resolved photoelectron spectroscopy (ARPES) which is a surface-sensitive technique and allows for the direct measurement of the $k$-resolved surface electronic structure. As a case study we present experiments for Cs atoms adsorbed on a Ag(111) substrate for coverages up to 0.06~monolayer (ML). For coverages above 0.03~ML we find the Fermi wave vector of the Ag(111) surface state to be increased by $\sim$100\% compared to the value for the clean surface. This remarkable observation gives a conclusive and quantitative explanation for the as yet surprisingly small lattice constant for the Cs superlattice on Ag(111).\cite{ziegler_scanning_2008} Our results thus disclose a new aspect of surface-state mediated ordering which can be essential for a proper understanding of the resulting adsorbate geometry. In turn, the mechanism provides new possibilities to tune electronic or magnetic interactions and corresponding ordering phenomena in low-dimensional nanostructures.

We performed high-resolution ARPES experiments employing a Scienta R4000 electron analyzer and a microwave-driven He-discharge lamp (MB-Scientific). The angular and energy resolution of the setup were set to 0.3$^\circ$ and 8~meV. For all presented measurements we used an excitation energy of $h\nu$~=~21.22~eV (He~I). The experiments were carried out under ultra-high vaccuum conditions with a base pressure below $2\cdot10^{-10}$ mbar and at a sample temperature of $\sim$20~K. The Ag(111) single-crystal was prepared by standard sputter-annealing cycles. The surface quality was characterized by ARPES measurements of the surface state line width (see Ref.~\onlinecite{reinert03}) and with low energy electron diffraction (LEED). Commercial alkali sources (SAES getters S.p.A.) were employed for Cs deposition. The deposition rate was gauged using the Cs-induced ($\sqrt{3}\mbox{x}\sqrt{3}$) reconstruction occuring at a coverage of 1/3~ML (see Ref.~\onlinecite{Diehl:96.2}) and is estimated to be accurate within $\pm$20~\%. 
\begin{figure}[t]
\includegraphics[width=3.2in]{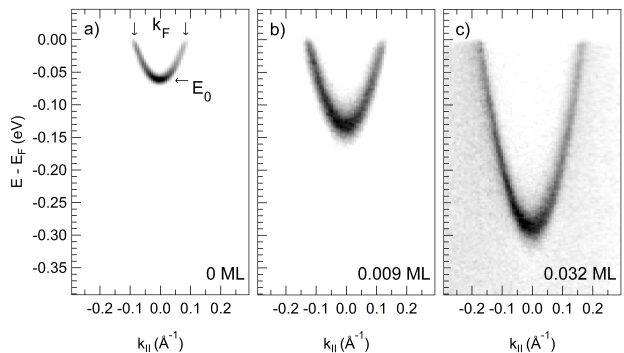}
\caption{\label{fig1} Angle-resolved photoemission for the clean (a) and Cs-covered (b)-(c) Ag(111) surface around the $\bar{\Gamma}$ point ($k_{||}$~=~0~\AA$^{-1}$). Fermi momentum $k_F$ and binding energy $E_0$ of the parabolic surface state dispersion are largely influenced by the Cs-induced modification of the surface potential barrier.}  
\end{figure}

The electronic structure of the pristine Ag(111) surface features a surface state in the bulk-projected energy gap around the $\bar{\Gamma}$-point of the surface Brillouin zone (Fig.~\ref{fig1}(a)). Its parabolic dispersion is specified by the Fermi wave vector $k_F$ and the maximum binding energy $E_0$ for which we determine values of 0.085~\AA$^{-1}$ and 62~meV, in agreement with previous experiments \cite{reinert03}. As inferred from Fig.~\ref{fig1}(b) and (c) successive deposition of impurity concentrations of Cs adatoms gives rise to a steady increase in the binding energy $E_0$ of the surface state. This pronounced dispersion modification results from the sensitivity of surface states to changes in the surface potential. When bound to a surface, alkali atoms tend to form comparably large dipole moments which results in significant reductions of the substrate work function \cite{bonzel:88}. Based on STM measurements on Cs/Ag(111) a work function change $\Delta\phi$~$\approx$~1eV has been determined for $\theta$~=~0.03-0.04~ML \cite{ziegler_scanning_2008}. Employing a simple phase analysis model to take into account the changed surface barrier (see Refs.~\onlinecite{Smith:85.9,echenique_theory_1989,Nuber:08.11}) and using this value for $\Delta\phi$ we estimate a surface state binding energy increase of $\sim$200~meV which is in good agreement with our experimental data in Fig.~\ref{fig1}(c). The large change in $E_0$ is therefore mainly attributed to modifications in the surface potential barrier due to the Cs adsorption whereas also direct doping most likely has an additional but smaller effect. We stress that the observation of a single well-defined surface state band for the Cs-covered samples indicates a homogeneous distribution of the Cs adatoms without island formation.\cite{Cercellier:06.5} This is indeed expected because of the dipole-dipole repulsion between the adsorbed atoms.
\begin{figure}[b]
\includegraphics[width=3.2in]{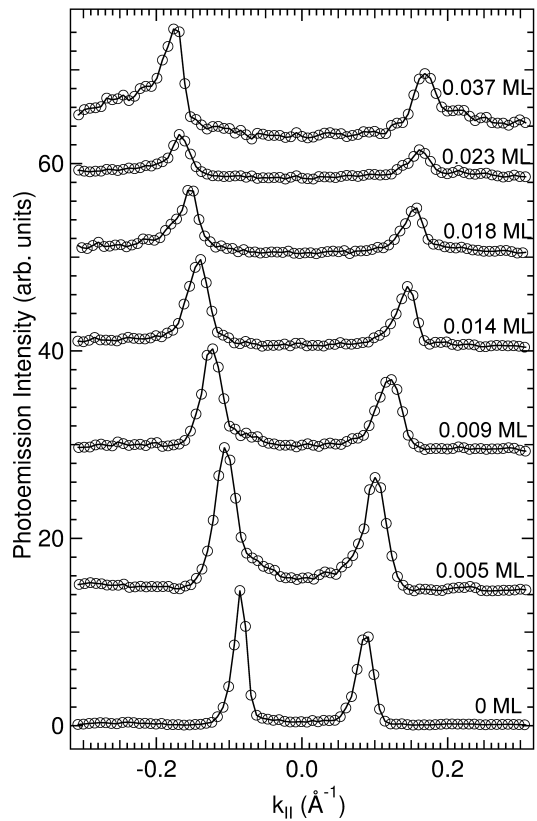}
\caption{\label{fig2}  Momentum distribution curves  at the Fermi energy $E_F$ for the Ag(111) surface state for increasing Cs coverage. The data evidence a coverage-dependent enhancement of the surface state Fermi wave vector $k_F$. The asymmetry in the peak intensities results from photoemission matrix element effects.}
\end{figure}

\begin{figure}[t]
\includegraphics[width=3.2in]{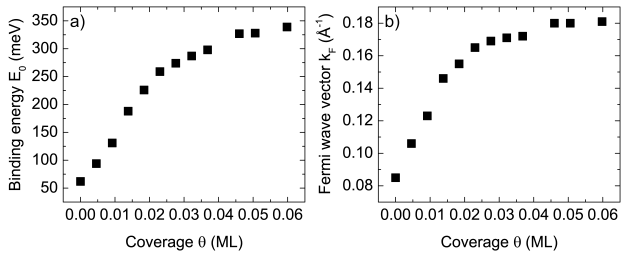}
\caption{\label{fig3} Full data set on the coverage dependence of the surface state parameters $E_0$ in (a) and $k_F$ in (b) corresponding to the representative data plots in Fig.~\ref{fig1} and Fig.~\ref{fig2}. The symbol size in (a) and (b) exceeds the experimental error along the ordinate axis.}
\end{figure}

A crucial consequence of the observed binding energy increase are associated changes of the surface state Fermi wave vector $k_F$. In Fig.~\ref{fig2} we consider momentum distribution curves (MDCs) taken at the Fermi energy $E_F$ for rising amounts of Cs coverage. The two peaks in each spectrum correspond to the Fermi wave vector in the $\pm k_{||}$ directions. The data indicate shifts of the two peaks to larger $k_{||}$ values and thus a gradual enhancement of the Fermi wave vector with increasing Cs concentration on the surface. The results of the full data set are summarized in  Fig.~\ref{fig3} where we plot the measured surface state parameters $k_F$ and $E_0$ as a function of the Cs coverage in panel (a) and (b). Both parameters show a similar coverage dependence: Up to $\sim$0.02~ML we find a linear increase which is maintained but reduced for higher coverages. The most important finding for the present work is that already at a Cs coverage of $\sim$0.03~ML the surface state Fermi wave vector is enhanced by a factor $>$2 and the binding energy is enhanced by a factor $>$4 when compared to values of the clean Ag(111) surface. It is important to note that the observed surface state modifications eventually result from chemisorptive, partially ionic Cs-Ag bonds with binding energies on the eV scale \cite{bonzel:88}. Therefore we do not expect a significant temperature dependence of the above findings in the low temperature regime below $\sim$100~K. 

In the following we will discuss the implications of the experimentally determined coverage dependence of the surface state parameters $k_F$ and $E_0$ for surface-state mediated interactions between impurities and a resulting superlattice formation. To this end we will compare our present results to previous STM experiments which observed the formation of a Cs adatom superlattice on Ag(111) at a coverage of 0.03-0.04~ML and a temperature of 7~K.\cite{ziegler_scanning_2008} The superlattice constant deduced in these experiments was $d_{Cs}$~=~(15$\pm$2)~{\AA} and hence considerably reduced as compared to the Ce superlattice on Ag(111) where $d_{Ce}$~=~(32$\pm$2)~{\AA} was found.\cite{silly_creation_2004}  

The oscillating interaction energy between two adatoms due to surface state scattering is given by
\begin{equation}
E^s_i = -A E_0 \left( \frac{2 \sin(\delta_0)}{\pi} \right)^2 \frac{\sin(2 k_F \vert r_1-r_2 \vert + 2 \delta_0)}{\left( k_F \vert r_1-r_2 \vert \right)^2 },
\end{equation}
with the scattering amplitude $A$ and the scattering phase $\delta_0$.\cite{Hyldgaard:00.1} It is important to note that the form of $E^s_i$ depends crucially on $k_F$ and $E_0$ and hence is expected to change significantly if these parameters are modified. In addition to $E^s_i$ the repulsive dipole-dipole interaction energy  
\begin{equation}
E^d_i = \frac{1}{4 \pi \epsilon_0} \frac{p^2}{\vert r_1-r_2 \vert^3}
\end{equation}
between two adatoms with the dipole moment $p$ must be taken into account to obtain the total interaction energy $E_i=E^s_i+E^d_i$. In Ref.~\onlinecite{ziegler_scanning_2008} the parameters $A$~$\approx$~0.3, $\delta_0$~$\approx$~0.5$\pi$ and $p$~$\approx$~(0.9$\pm$0.3)~e$\cdot${\AA} were estimated for Cs/Ag(111). Using these values we are able to compute the interaction energy $E_i$ between two Cs atoms on the Ag(111) surface for the measured parameter pairs of $k_F$ and $E_0$ (Fig.~\ref{fig3}).
\begin{figure}[b]
\includegraphics[width=3.2in]{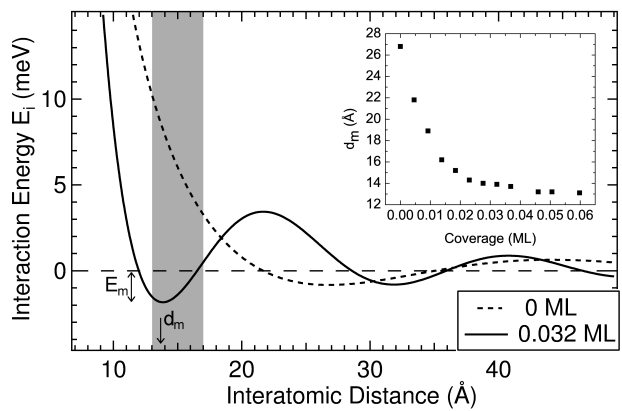}
\caption{\label{fig4} Calculated adatom interaction energy $E_i$ using surface state parameters of the clean surface ($k_F$~=~0.085~~\AA$^{-1}$, $E_0$~=~62~meV; dashed line) and of the Cs-covered surface for $\theta$~=~0.032~ML ($k_F$~=~0.171~~\AA$^{-1}$, $E_0$~=~287~meV; full line). The shaded area represents the Cs superlattice constant including experimental error found by STM experiments \cite{ziegler_scanning_2008}. The inset shows the calculated position $d_m$ of first energy minimum $E_m$ (see text for details) as a function of the Cs coverage $\theta$. All calculations are done for $\delta_0$~=~0.05$\pi$.}
\end{figure}

In Fig.~\ref{fig4} we plot $E_i$ for the surface state parameters of the clean surface (dashed line) and for the ones obtained by our measurements for $\theta$~=~0.032~ML (full line). The most important parameter in these curves is the location of the first energy minimum $d_{m}$ as for this interatomic distance the formation of an energetically stabilized superlattice is possible. Clearly, for the parameters of the clean surface $d_{m}$ largely deviates from the lattice constant of 15~{\AA} found in Ref.~\onlinecite{ziegler_scanning_2008} (see shaded area in Fig.~\ref{fig4}). However, using the re-determined values from our ARPES study the agreement is striking. We find $d_{m}\approx$~14~{\AA} which fully coincides with the previously observed superlattice constant within experimental uncertainty. The large change in $d_{m}$ results from the modification of the Fermi wave vector after Cs deposition. Note that the distance $d_{m}$ does not only depend on $k_F$ but also on the scattering phase $\delta_0$ the choice of which constituting the largest uncertainty in the present determination of $d_{m}$. Previous investigations found scattering phases of 0.33$\pi$ and 0.37$\pi$ for Co and Ce atoms on Ag(111).\cite{knorr_long-range_2002,silly_creation_2004} Assuming according reductions in $\delta_0$ we obtain $d_{m}\approx$~16-17~{\AA}. The agreement with the STM experiments is thus maintained for a plausible range of scattering phases whereas optimal correspondence ($d_{m}\approx$~15~{\AA}) is found for $\delta_0$~=~0.43$\pi$.       

Importantly, in addition to the modified Fermi wave vector also the strong increase of the binding energy $E_0$ plays an essential role for our considerations. $E_0$ defines the energy scale in Eq.~(1) and therefore determines the actual energy gain $E_{m}$ for a given scattering phase and amplitude (see Fig.~\ref{fig4}). The importance of a sufficiently large ratio between $E_m$ and the adatom diffusion barrier height $E_D$ for the realization of a superlattice has recently been emphasized.\cite{Kirschner:10.3} For the system Cs/Ag(111) $E_D$~$\approx$~17~meV was determined in Ref.~\onlinecite{ziegler_scanning_2008}. Using this number we obtain values for $E_{m}$/$E_D$ in the range of $\sim$7-12~\% for scattering phases $\delta_0$ between 0.37$\pi$ and 0.5$\pi$. Note here that $E_{m}$, unlike $d_{m}$, additionally shows a strong dependence on the scattering amplitude $A$ and the dipole moment $p$ rendering our calculated values as a rough estimation. Nevertheless, the numbers we obtain agree with those observed for other adsorbate-substrate combinations which were shown to stabilize adatom superlattices where one typically finds $E_{m}$/$E_D$~$>$~6\%.\cite{Kirschner:10.3} The strong enhancement of $E_0$ with Cs adsorption can therefore be indentified as an effect of critical importance for the energetical stabilization of the superlattice. 

Taken collectively our experimental results provide a comprehensive explanation for the observation of a Cs superlattice on Ag(111) with a lattice constant of 15~{\AA} and a coverage of 0.03-0.04~ML. We deduce that the structural ordering in this system is not solely determined by the substrate but additionally by the adsorbate itself which decisively influences the surface-state mediated interaction. It is instructive to examine the coverage dependence of the equilibrium distance $d_m$ obtained from the data in Fig.~\ref{fig3} (see inset in Fig.~\ref{fig4}). It turns out that the energy minimum distance of the adatom interaction energy varies over a considerable range between 13~{\AA} and 27~{\AA}. This demonstrates a new possibility to tune surface-state mediated interactions between surface adatoms, in addition to a change of the substrate.\cite{knorr_long-range_2002}

As several previous studies on different noble metals reported surface state modifications for a variety of different adsorbate species, the findings presented here are likely to have consequences for a broad range of material systems [see e.g. Ref.~\onlinecite{forster_systematic_2006}]. Generally, electropositive species, such as alkali atoms studied here, tend increase the surface state binding energy $E_0$ and Fermi wave vector $k_F$.\cite{Lindgren:78} However, also the opposite effect, that is a reduction of $E_0$ and $k_F$, has been observed, for example for the adsorbtion of CO molecules\cite{Lindgren:82}, rare gases\cite{Reihl:01} or dielectric NaCl overlayers\cite{Repp:04.1} on noble metal surfaces. In this case an enhancement of the equilibrium distance $d_m$ should be expected. For possible implications of our work we point out to a number of recent studies on atomic or molecular superstructures on metal surfaces for which a significant change in the surface state dispersion due to dipole formations may be expected.\cite{Ternes:10.1,Yokoyama:07.5,Pascual:07.10} Interestingly, another investigation reports variations in the distance distributions for different organic and inorganic molecules on Cu(111) depending on their dipole moments.\cite{morgenstern:10.7} One can anticipate that the mechanism established in our work plays an important role for these observations. Furthermore, adsorption experiments on topological insulators find modifications in the surface electronic structure upon adatom coverage suggesting that similar effects as discussed here will also be relevant for the surface-state-mediated RKKY-interactions and resulting surface magnetism predicted for these systems.\cite{Hasan:11.1,Zhang:09.4,Chang:11.2}   

We have established a new mechanism in the adsorbate-substrate interplay in low-dimensional surface nanostructures which, as we have shown, can radically change the involved interactions between adsorbates. In particular, surface-state mediated interactions are found to be modified upon adsorbate deposition. This is due to the sensitivity of surface states to changes in the vacuum-surface potential barrier. The findings offer a new way to manipulate the self-assembly of atoms or molecules on surfaces.  

We acknowledge helpful discussions with Matthias Bode, Frank Forster and Mattia Mulazzi. This work was supported by the Bundesministerium f\"ur Bildung und Forschung (Grants No. 05K10WW1/2 and No. 05KS1WMB/1) and the Deutsche Forschunsgsgemeinschaft within the Forschergruppe 1162.
\bibliographystyle{apsrev}

\begin{thebibliography}{40}
\expandafter\ifx\csname natexlab\endcsname\relax\def\natexlab#1{#1}\fi
\expandafter\ifx\csname bibnamefont\endcsname\relax
  \def\bibnamefont#1{#1}\fi
\expandafter\ifx\csname bibfnamefont\endcsname\relax
  \def\bibfnamefont#1{#1}\fi
\expandafter\ifx\csname citenamefont\endcsname\relax
  \def\citenamefont#1{#1}\fi
\expandafter\ifx\csname url\endcsname\relax
  \def\url#1{\texttt{#1}}\fi
\expandafter\ifx\csname urlprefix\endcsname\relax\def\urlprefix{URL }\fi
\providecommand{\bibinfo}[2]{#2}
\providecommand{\eprint}[2][]{\url{#2}}

\bibitem[{\citenamefont{Park et~al.}(2007)\citenamefont{Park, Lim, Lee, and
  Cho}}]{Cho:07}
\bibinfo{author}{\bibfnamefont{Y.~D.} \bibnamefont{Park}},
  \bibinfo{author}{\bibfnamefont{J.~A.} \bibnamefont{Lim}},
  \bibinfo{author}{\bibfnamefont{H.~S.} \bibnamefont{Lee}}, \bibnamefont{and}
  \bibinfo{author}{\bibfnamefont{K.}~\bibnamefont{Cho}},
  \bibinfo{journal}{Mater. Today} \textbf{\bibinfo{volume}{10}},
  \bibinfo{pages}{46} (\bibinfo{year}{2007}).

\bibitem[{\citenamefont{Blumenstein et~al.}(2011)\citenamefont{Blumenstein,
  Sch\"afer, Mietke, Meyer, Dollinger, Lochner, Cui, Patthey, Matzdorf, and
  Claessen}}]{Schaefer:11.10}
\bibinfo{author}{\bibfnamefont{C.}~\bibnamefont{Blumenstein}},
  \bibinfo{author}{\bibfnamefont{J.}~\bibnamefont{Sch\"afer}},
  \bibinfo{author}{\bibfnamefont{S.}~\bibnamefont{Mietke}},
  \bibinfo{author}{\bibfnamefont{S.}~\bibnamefont{Meyer}},
  \bibinfo{author}{\bibfnamefont{A.}~\bibnamefont{Dollinger}},
  \bibinfo{author}{\bibfnamefont{M.}~\bibnamefont{Lochner}},
  \bibinfo{author}{\bibfnamefont{X.~Y.} \bibnamefont{Cui}},
  \bibinfo{author}{\bibfnamefont{L.}~\bibnamefont{Patthey}},
  \bibinfo{author}{\bibfnamefont{R.}~\bibnamefont{Matzdorf}}, \bibnamefont{and}
  \bibinfo{author}{\bibfnamefont{R.}~\bibnamefont{Claessen}},
  \bibinfo{journal}{Nat Phys} \textbf{\bibinfo{volume}{7}},
  \bibinfo{pages}{776} (\bibinfo{year}{2011}).

\bibitem[{\citenamefont{Bode et~al.}(2007)\citenamefont{Bode, Heide, von
  Bergmann, Ferriani, Heinze, Bihlmayer, Kubetzka, Pietzsch, Bl\"ugel, and
  Wiesendanger}}]{Bode:07.5}
\bibinfo{author}{\bibfnamefont{M.}~\bibnamefont{Bode}},
  \bibinfo{author}{\bibfnamefont{M.}~\bibnamefont{Heide}},
  \bibinfo{author}{\bibfnamefont{K.}~\bibnamefont{von Bergmann}},
  \bibinfo{author}{\bibfnamefont{P.}~\bibnamefont{Ferriani}},
  \bibinfo{author}{\bibfnamefont{S.}~\bibnamefont{Heinze}},
  \bibinfo{author}{\bibfnamefont{G.}~\bibnamefont{Bihlmayer}},
  \bibinfo{author}{\bibfnamefont{A.}~\bibnamefont{Kubetzka}},
  \bibinfo{author}{\bibfnamefont{O.}~\bibnamefont{Pietzsch}},
  \bibinfo{author}{\bibfnamefont{S.}~\bibnamefont{Bl\"ugel}}, \bibnamefont{and}
  \bibinfo{author}{\bibfnamefont{R.}~\bibnamefont{Wiesendanger}},
  \bibinfo{journal}{Nature} \textbf{\bibinfo{volume}{447}},
  \bibinfo{pages}{190} (\bibinfo{year}{2007}).

\bibitem[{\citenamefont{Gambardella et~al.}(2009)\citenamefont{Gambardella,
  Stepanow, Dmitriev, Honolka, de~Groot, Lingenfelder, Gupta, Sarma, Bencok,
  Stanescu et~al.}}]{gambardella:09.3}
\bibinfo{author}{\bibfnamefont{P.}~\bibnamefont{Gambardella}},
  \bibinfo{author}{\bibfnamefont{S.}~\bibnamefont{Stepanow}},
  \bibinfo{author}{\bibfnamefont{A.}~\bibnamefont{Dmitriev}},
  \bibinfo{author}{\bibfnamefont{J.}~\bibnamefont{Honolka}},
  \bibinfo{author}{\bibfnamefont{F.~M.~F.} \bibnamefont{de~Groot}},
  \bibinfo{author}{\bibfnamefont{M.}~\bibnamefont{Lingenfelder}},
  \bibinfo{author}{\bibfnamefont{S.~S.} \bibnamefont{Gupta}},
  \bibinfo{author}{\bibfnamefont{D.~D.} \bibnamefont{Sarma}},
  \bibinfo{author}{\bibfnamefont{P.}~\bibnamefont{Bencok}},
  \bibinfo{author}{\bibfnamefont{S.}~\bibnamefont{Stanescu}},
  \bibnamefont{et~al.}, \bibinfo{journal}{Nat Mater}
  \textbf{\bibinfo{volume}{8}}, \bibinfo{pages}{189} (\bibinfo{year}{2009}).

\bibitem[{\citenamefont{Sakamoto et~al.}(2009)\citenamefont{Sakamoto, Oda,
  Kimura, Miyamoto, Tsujikawa, Imai, Ueno, Namatame, Taniguchi, Eriksson
  et~al.}}]{Sakamoto:09.3}
\bibinfo{author}{\bibfnamefont{K.}~\bibnamefont{Sakamoto}},
  \bibinfo{author}{\bibfnamefont{T.}~\bibnamefont{Oda}},
  \bibinfo{author}{\bibfnamefont{A.}~\bibnamefont{Kimura}},
  \bibinfo{author}{\bibfnamefont{K.}~\bibnamefont{Miyamoto}},
  \bibinfo{author}{\bibfnamefont{M.}~\bibnamefont{Tsujikawa}},
  \bibinfo{author}{\bibfnamefont{A.}~\bibnamefont{Imai}},
  \bibinfo{author}{\bibfnamefont{N.}~\bibnamefont{Ueno}},
  \bibinfo{author}{\bibfnamefont{H.}~\bibnamefont{Namatame}},
  \bibinfo{author}{\bibfnamefont{M.}~\bibnamefont{Taniguchi}},
  \bibinfo{author}{\bibfnamefont{P.~E.~J.} \bibnamefont{Eriksson}},
  \bibnamefont{et~al.}, \bibinfo{journal}{Phys. Rev. Lett.}
  \textbf{\bibinfo{volume}{102}}, \bibinfo{pages}{096805}
  (\bibinfo{year}{2009}).

\bibitem[{\citenamefont{Abd El-Fattah et~al.}(2011)\citenamefont{Abd El-Fattah,
  Matena, Corso, Garcia~de Abajo, Schiller, and Ortega}}]{Ortega:11.8}
\bibinfo{author}{\bibfnamefont{Z.~M.} \bibnamefont{Abd El-Fattah}},
  \bibinfo{author}{\bibfnamefont{M.}~\bibnamefont{Matena}},
  \bibinfo{author}{\bibfnamefont{M.}~\bibnamefont{Corso}},
  \bibinfo{author}{\bibfnamefont{F.~J.} \bibnamefont{Garcia~de Abajo}},
  \bibinfo{author}{\bibfnamefont{F.}~\bibnamefont{Schiller}}, \bibnamefont{and}
  \bibinfo{author}{\bibfnamefont{J.~E.} \bibnamefont{Ortega}},
  \bibinfo{journal}{Phys. Rev. Lett.} \textbf{\bibinfo{volume}{107}},
  \bibinfo{pages}{066803} (\bibinfo{year}{2011}).

\bibitem[{\citenamefont{Wang et~al.}(2009)\citenamefont{Wang, Ge, Manzano,
  Kr\"oger, Berndt, Hofer, Tang, and Cerda}}]{Berndt:09}
\bibinfo{author}{\bibfnamefont{Y.}~\bibnamefont{Wang}},
  \bibinfo{author}{\bibfnamefont{X.}~\bibnamefont{Ge}},
  \bibinfo{author}{\bibfnamefont{C.}~\bibnamefont{Manzano}},
  \bibinfo{author}{\bibfnamefont{J.}~\bibnamefont{Kr\"oger}},
  \bibinfo{author}{\bibfnamefont{R.}~\bibnamefont{Berndt}},
  \bibinfo{author}{\bibfnamefont{W.~A.} \bibnamefont{Hofer}},
  \bibinfo{author}{\bibfnamefont{H.}~\bibnamefont{Tang}}, \bibnamefont{and}
  \bibinfo{author}{\bibfnamefont{J.}~\bibnamefont{Cerda}}, \bibinfo{journal}{J.
  Am. Chem. Soc.} \textbf{\bibinfo{volume}{131}}, \bibinfo{pages}{10400}
  (\bibinfo{year}{2009}).

\bibitem[{\citenamefont{Stadler et~al.}(2009)\citenamefont{Stadler, Hansen,
  Kr\"oger, Kumpf, and Umbach}}]{Kumpf:09.2}
\bibinfo{author}{\bibfnamefont{C.}~\bibnamefont{Stadler}},
  \bibinfo{author}{\bibfnamefont{S.}~\bibnamefont{Hansen}},
  \bibinfo{author}{\bibfnamefont{I.}~\bibnamefont{Kr\"oger}},
  \bibinfo{author}{\bibfnamefont{C.}~\bibnamefont{Kumpf}}, \bibnamefont{and}
  \bibinfo{author}{\bibfnamefont{E.}~\bibnamefont{Umbach}},
  \bibinfo{journal}{Nat Phys} \textbf{\bibinfo{volume}{5}},
  \bibinfo{pages}{153} (\bibinfo{year}{2009}).

\bibitem[{\citenamefont{Otero et~al.}(2011)\citenamefont{Otero, Gallego,
  de~Parga, Martin, and Miranda}}]{otero:11.11}
\bibinfo{author}{\bibfnamefont{R.}~\bibnamefont{Otero}},
  \bibinfo{author}{\bibfnamefont{J.~M.} \bibnamefont{Gallego}},
  \bibinfo{author}{\bibfnamefont{A.~L.~V.} \bibnamefont{de~Parga}},
  \bibinfo{author}{\bibfnamefont{N.}~\bibnamefont{Martin}}, \bibnamefont{and}
  \bibinfo{author}{\bibfnamefont{R.}~\bibnamefont{Miranda}},
  \bibinfo{journal}{Advanced Materials} \textbf{\bibinfo{volume}{23}},
  \bibinfo{pages}{5148} (\bibinfo{year}{2011}).

\bibitem[{\citenamefont{Barth et~al.}(2005)\citenamefont{Barth, Costantini, and
  Kern}}]{barth:05}
\bibinfo{author}{\bibfnamefont{J.~V.} \bibnamefont{Barth}},
  \bibinfo{author}{\bibfnamefont{G.}~\bibnamefont{Costantini}},
  \bibnamefont{and} \bibinfo{author}{\bibfnamefont{K.}~\bibnamefont{Kern}},
  \bibinfo{journal}{Nature} \textbf{\bibinfo{volume}{437}},
  \bibinfo{pages}{671} (\bibinfo{year}{2005}).

\bibitem[{\citenamefont{Forster et~al.}(2006)\citenamefont{Forster, Bendounan,
  Ziroff, and Reinert}}]{forster_systematic_2006}
\bibinfo{author}{\bibfnamefont{F.}~\bibnamefont{Forster}},
  \bibinfo{author}{\bibfnamefont{A.}~\bibnamefont{Bendounan}},
  \bibinfo{author}{\bibfnamefont{J.}~\bibnamefont{Ziroff}}, \bibnamefont{and}
  \bibinfo{author}{\bibfnamefont{F.}~\bibnamefont{Reinert}},
  \bibinfo{journal}{Surface Science} \textbf{\bibinfo{volume}{600}},
  \bibinfo{pages}{3870} (\bibinfo{year}{2006}).

\bibitem[{\citenamefont{Han and Weiss}(2012)}]{Weiss:12.01}
\bibinfo{author}{\bibfnamefont{P.}~\bibnamefont{Han}} \bibnamefont{and}
  \bibinfo{author}{\bibfnamefont{P.~S.} \bibnamefont{Weiss}},
  \bibinfo{journal}{Surface Science Reports} \textbf{\bibinfo{volume}{67}},
  \bibinfo{pages}{19 } (\bibinfo{year}{2012}).

\bibitem[{\citenamefont{Repp et~al.}(2000)\citenamefont{Repp, Moresco, Meyer,
  Rieder, Hyldgaard, and Persson}}]{repp_substrate_2000}
\bibinfo{author}{\bibfnamefont{J.}~\bibnamefont{Repp}},
  \bibinfo{author}{\bibfnamefont{F.}~\bibnamefont{Moresco}},
  \bibinfo{author}{\bibfnamefont{G.}~\bibnamefont{Meyer}},
  \bibinfo{author}{\bibfnamefont{K.H.}~\bibnamefont{Rieder}},
  \bibinfo{author}{\bibfnamefont{P.}~\bibnamefont{Hyldgaard}},
  \bibnamefont{and} \bibinfo{author}{\bibfnamefont{M.}~\bibnamefont{Persson}},
  \bibinfo{journal}{Physical Review Letters} \textbf{\bibinfo{volume}{85}},
  \bibinfo{pages}{2981} (\bibinfo{year}{2000}).

\bibitem[{\citenamefont{Knorr et~al.}(2002)\citenamefont{Knorr, Brune, Epple,
  Hirstein, Schneider, and Kern}}]{knorr_long-range_2002}
\bibinfo{author}{\bibfnamefont{N.}~\bibnamefont{Knorr}},
  \bibinfo{author}{\bibfnamefont{H.}~\bibnamefont{Brune}},
  \bibinfo{author}{\bibfnamefont{M.}~\bibnamefont{Epple}},
  \bibinfo{author}{\bibfnamefont{A.}~\bibnamefont{Hirstein}},
  \bibinfo{author}{\bibfnamefont{M.~A.} \bibnamefont{Schneider}},
  \bibnamefont{and} \bibinfo{author}{\bibfnamefont{K.}~\bibnamefont{Kern}},
  \bibinfo{journal}{Physical Review B} \textbf{\bibinfo{volume}{65}},
  \bibinfo{pages}{115420} (\bibinfo{year}{2002}).

\bibitem[{\citenamefont{Liu et~al.}(2009)\citenamefont{Liu, Liu, Xu, Qi, and
  Zhang}}]{Zhang:09.4}
\bibinfo{author}{\bibfnamefont{Q.}~\bibnamefont{Liu}},
  \bibinfo{author}{\bibfnamefont{C.-X.} \bibnamefont{Liu}},
  \bibinfo{author}{\bibfnamefont{C.}~\bibnamefont{Xu}},
  \bibinfo{author}{\bibfnamefont{X.-L.} \bibnamefont{Qi}}, \bibnamefont{and}
  \bibinfo{author}{\bibfnamefont{S.-C.} \bibnamefont{Zhang}},
  \bibinfo{journal}{Phys. Rev. Lett.} \textbf{\bibinfo{volume}{102}},
  \bibinfo{pages}{156603} (\bibinfo{year}{2009}).

\bibitem[{\citenamefont{Stepanyuk et~al.}(2004)\citenamefont{Stepanyuk,
  Niebergall, Longo, Hergert, and Bruno}}]{Stepanyuk:04.8}
\bibinfo{author}{\bibfnamefont{V.~S.} \bibnamefont{Stepanyuk}},
  \bibinfo{author}{\bibfnamefont{L.}~\bibnamefont{Niebergall}},
  \bibinfo{author}{\bibfnamefont{R.~C.} \bibnamefont{Longo}},
  \bibinfo{author}{\bibfnamefont{W.}~\bibnamefont{Hergert}}, \bibnamefont{and}
  \bibinfo{author}{\bibfnamefont{P.}~\bibnamefont{Bruno}},
  \bibinfo{journal}{Phys. Rev. B} \textbf{\bibinfo{volume}{70}},
  \bibinfo{pages}{075414} (\bibinfo{year}{2004}).

\bibitem[{\citenamefont{Crommie et~al.}(1993)\citenamefont{Crommie, Lutz, and
  Eigler}}]{crommie_imaging_1993}
\bibinfo{author}{\bibfnamefont{M.~F.} \bibnamefont{Crommie}},
  \bibinfo{author}{\bibfnamefont{C.~P.} \bibnamefont{Lutz}}, \bibnamefont{and}
  \bibinfo{author}{\bibfnamefont{D.~M.} \bibnamefont{Eigler}},
  \bibinfo{journal}{Nature} \textbf{\bibinfo{volume}{363}},
  \bibinfo{pages}{524} (\bibinfo{year}{1993}).

\bibitem[{\citenamefont{Hyldgaard and Persson}(2000)}]{Hyldgaard:00.1}
\bibinfo{author}{\bibfnamefont{P.}~\bibnamefont{Hyldgaard}} \bibnamefont{and}
  \bibinfo{author}{\bibfnamefont{M.}~\bibnamefont{Persson}},
  \bibinfo{journal}{Journal of Physics: Condensed Matter}
  \textbf{\bibinfo{volume}{12}}, \bibinfo{pages}{L13} (\bibinfo{year}{2000}).

\bibitem[{\citenamefont{Ternes et~al.}(2010)\citenamefont{Ternes, Pivetta,
  Patthey, and Schneider}}]{Ternes:10.1}
\bibinfo{author}{\bibfnamefont{M.}~\bibnamefont{Ternes}},
  \bibinfo{author}{\bibfnamefont{M.}~\bibnamefont{Pivetta}},
  \bibinfo{author}{\bibfnamefont{F.}~\bibnamefont{Patthey}}, \bibnamefont{and}
  \bibinfo{author}{\bibfnamefont{W.D.}~\bibnamefont{Schneider}},
  \bibinfo{journal}{Progress in Surface Science} \textbf{\bibinfo{volume}{85}},
  \bibinfo{pages}{1} (\bibinfo{year}{2010}).

\bibitem[{\citenamefont{Silly et~al.}(2004)\citenamefont{Silly, Pivetta,
  Ternes, Patthey, Pelz, and Schneider}}]{silly_creation_2004}
\bibinfo{author}{\bibfnamefont{F.}~\bibnamefont{Silly}},
  \bibinfo{author}{\bibfnamefont{M.}~\bibnamefont{Pivetta}},
  \bibinfo{author}{\bibfnamefont{M.}~\bibnamefont{Ternes}},
  \bibinfo{author}{\bibfnamefont{F.}~\bibnamefont{Patthey}},
  \bibinfo{author}{\bibfnamefont{J.~P.} \bibnamefont{Pelz}}, \bibnamefont{and}
  \bibinfo{author}{\bibfnamefont{W.D.}~\bibnamefont{Schneider}},
  \bibinfo{journal}{Physical Review Letters} \textbf{\bibinfo{volume}{92}},
  \bibinfo{pages}{016101} (\bibinfo{year}{2004}).

\bibitem[{\citenamefont{Negulyaev et~al.}(2009)\citenamefont{Negulyaev,
  Stepanyuk, Niebergall, Bruno, Pivetta, Ternes, Patthey, and
  Schneider}}]{Schneider:09.6}
\bibinfo{author}{\bibfnamefont{N.~N.} \bibnamefont{Negulyaev}},
  \bibinfo{author}{\bibfnamefont{V.~S.} \bibnamefont{Stepanyuk}},
  \bibinfo{author}{\bibfnamefont{L.}~\bibnamefont{Niebergall}},
  \bibinfo{author}{\bibfnamefont{P.}~\bibnamefont{Bruno}},
  \bibinfo{author}{\bibfnamefont{M.}~\bibnamefont{Pivetta}},
  \bibinfo{author}{\bibfnamefont{M.}~\bibnamefont{Ternes}},
  \bibinfo{author}{\bibfnamefont{F.}~\bibnamefont{Patthey}}, \bibnamefont{and}
  \bibinfo{author}{\bibfnamefont{W.-D.} \bibnamefont{Schneider}},
  \bibinfo{journal}{Phys. Rev. Lett.} \textbf{\bibinfo{volume}{102}},
  \bibinfo{pages}{246102} (\bibinfo{year}{2009}).

\bibitem[{\citenamefont{von Hofe et~al.}(2006)\citenamefont{von Hofe, Kr\"oger,
  and Berndt}}]{Berndt:06.7}
\bibinfo{author}{\bibfnamefont{T.}~\bibnamefont{von Hofe}},
  \bibinfo{author}{\bibfnamefont{J.}~\bibnamefont{Kr\"oger}}, \bibnamefont{and}
  \bibinfo{author}{\bibfnamefont{R.}~\bibnamefont{Berndt}},
  \bibinfo{journal}{Phys. Rev. B} \textbf{\bibinfo{volume}{73}},
  \bibinfo{pages}{245434} (\bibinfo{year}{2006}).

\bibitem[{\citenamefont{Ziegler et~al.}(2008)\citenamefont{Ziegler, Kr\"oger,
  Berndt, Filinov, and Bonitz}}]{ziegler_scanning_2008}
\bibinfo{author}{\bibfnamefont{M.}~\bibnamefont{Ziegler}},
  \bibinfo{author}{\bibfnamefont{J.}~\bibnamefont{Kr\"oger}},
  \bibinfo{author}{\bibfnamefont{R.}~\bibnamefont{Berndt}},
  \bibinfo{author}{\bibfnamefont{A.}~\bibnamefont{Filinov}}, \bibnamefont{and}
  \bibinfo{author}{\bibfnamefont{M.}~\bibnamefont{Bonitz}},
  \bibinfo{journal}{Physical Review B} \textbf{\bibinfo{volume}{78}},
  \bibinfo{pages}{245427} (\bibinfo{year}{2008}).

\bibitem[{\citenamefont{Reinert et~al.}(2001)\citenamefont{Reinert, Nicolay,
  Schmidt, Ehm, and H\"ufner}}]{reinert03}
\bibinfo{author}{\bibfnamefont{F.}~\bibnamefont{Reinert}},
  \bibinfo{author}{\bibfnamefont{G.}~\bibnamefont{Nicolay}},
  \bibinfo{author}{\bibfnamefont{S.}~\bibnamefont{Schmidt}},
  \bibinfo{author}{\bibfnamefont{D.}~\bibnamefont{Ehm}}, \bibnamefont{and}
  \bibinfo{author}{\bibfnamefont{S.}~\bibnamefont{H\"ufner}},
  \bibinfo{journal}{Phys. Rev. B} \textbf{\bibinfo{volume}{63}},
  \bibinfo{pages}{115415} (\bibinfo{year}{2001}).

\bibitem[{\citenamefont{Leatherman and Diehl}(1996)}]{Diehl:96.2}
\bibinfo{author}{\bibfnamefont{G.~S.} \bibnamefont{Leatherman}}
  \bibnamefont{and} \bibinfo{author}{\bibfnamefont{R.~D.} \bibnamefont{Diehl}},
  \bibinfo{journal}{Phys. Rev. B} \textbf{\bibinfo{volume}{53}},
  \bibinfo{pages}{4939} (\bibinfo{year}{1996}).

\bibitem[{\citenamefont{Bonzel}(1988)}]{bonzel:88}
\bibinfo{author}{\bibfnamefont{H.}~\bibnamefont{Bonzel}},
  \bibinfo{journal}{Surface Science Reports} \textbf{\bibinfo{volume}{8}},
  \bibinfo{pages}{43} (\bibinfo{year}{1988}).

\bibitem[{\citenamefont{Echenique and Pendry}(1989)}]{echenique_theory_1989}
\bibinfo{author}{\bibfnamefont{P.}~\bibnamefont{Echenique}} \bibnamefont{and}
  \bibinfo{author}{\bibfnamefont{J.}~\bibnamefont{Pendry}},
  \bibinfo{journal}{Progress in Surface Science} \textbf{\bibinfo{volume}{32}},
  \bibinfo{pages}{111} (\bibinfo{year}{1989}).

\bibitem[{\citenamefont{Nuber et~al.}(2008)\citenamefont{Nuber, Higashiguchi,
  Forster, Blaha, Shimada, and Reinert}}]{Nuber:08.11}
\bibinfo{author}{\bibfnamefont{A.}~\bibnamefont{Nuber}},
  \bibinfo{author}{\bibfnamefont{M.}~\bibnamefont{Higashiguchi}},
  \bibinfo{author}{\bibfnamefont{F.}~\bibnamefont{Forster}},
  \bibinfo{author}{\bibfnamefont{P.}~\bibnamefont{Blaha}},
  \bibinfo{author}{\bibfnamefont{K.}~\bibnamefont{Shimada}}, \bibnamefont{and}
  \bibinfo{author}{\bibfnamefont{F.}~\bibnamefont{Reinert}},
  \bibinfo{journal}{Phys. Rev. B} \textbf{\bibinfo{volume}{78}},
  \bibinfo{pages}{195412} (\bibinfo{year}{2008}).

\bibitem[{\citenamefont{Smith}(1985)}]{Smith:85.9}
\bibinfo{author}{\bibfnamefont{N.~V.} \bibnamefont{Smith}},
  \bibinfo{journal}{Phys. Rev. B} \textbf{\bibinfo{volume}{32}},
  \bibinfo{pages}{3549} (\bibinfo{year}{1985}).

\bibitem[{\citenamefont{Cercellier et~al.}(2006)\citenamefont{Cercellier,
  Didiot, Fagot-Revurat, Kierren, Moreau, Malterre, and
  Reinert}}]{Cercellier:06.5}
\bibinfo{author}{\bibfnamefont{H.}~\bibnamefont{Cercellier}},
  \bibinfo{author}{\bibfnamefont{C.}~\bibnamefont{Didiot}},
  \bibinfo{author}{\bibfnamefont{Y.}~\bibnamefont{Fagot-Revurat}},
  \bibinfo{author}{\bibfnamefont{B.}~\bibnamefont{Kierren}},
  \bibinfo{author}{\bibfnamefont{L.}~\bibnamefont{Moreau}},
  \bibinfo{author}{\bibfnamefont{D.}~\bibnamefont{Malterre}}, \bibnamefont{and}
  \bibinfo{author}{\bibfnamefont{F.}~\bibnamefont{Reinert}},
  \bibinfo{journal}{Phys. Rev. B} \textbf{\bibinfo{volume}{73}},
  \bibinfo{pages}{195413} (\bibinfo{year}{2006}).

\bibitem[{\citenamefont{Zhang et~al.}(2010)\citenamefont{Zhang, Miao, Sun, Gao,
  Hu, Ding, and Kirschner}}]{Kirschner:10.3}
\bibinfo{author}{\bibfnamefont{X.~P.} \bibnamefont{Zhang}},
  \bibinfo{author}{\bibfnamefont{B.~F.} \bibnamefont{Miao}},
  \bibinfo{author}{\bibfnamefont{L.}~\bibnamefont{Sun}},
  \bibinfo{author}{\bibfnamefont{C.~L.} \bibnamefont{Gao}},
  \bibinfo{author}{\bibfnamefont{A.}~\bibnamefont{Hu}},
  \bibinfo{author}{\bibfnamefont{H.~F.} \bibnamefont{Ding}}, \bibnamefont{and}
  \bibinfo{author}{\bibfnamefont{J.}~\bibnamefont{Kirschner}},
  \bibinfo{journal}{Phys. Rev. B} \textbf{\bibinfo{volume}{81}},
  \bibinfo{pages}{125438} (\bibinfo{year}{2010}).

\bibitem[{\citenamefont{Lindgren and Walld\'en}(1978)}]{Lindgren:78}
\bibinfo{author}{\bibfnamefont{S.~{\AA}.}~\bibnamefont{Lindgren}} \bibnamefont{and}
  \bibinfo{author}{\bibfnamefont{L.}~\bibnamefont{Walld\'en}},
  \bibinfo{journal}{Solid State Communications} \textbf{\bibinfo{volume}{28}},
  \bibinfo{pages}{283 } (\bibinfo{year}{1978}).

\bibitem[{\citenamefont{Lindgren et~al.}(1982)\citenamefont{Lindgren, Paul, and
  Walld\'en}}]{Lindgren:82}
\bibinfo{author}{\bibfnamefont{S.~{\AA}.}~\bibnamefont{Lindgren}},
  \bibinfo{author}{\bibfnamefont{J.}~\bibnamefont{Paul}}, \bibnamefont{and}
  \bibinfo{author}{\bibfnamefont{L.}~\bibnamefont{Walld\'en}},
  \bibinfo{journal}{Surface Science} \textbf{\bibinfo{volume}{117}},
  \bibinfo{pages}{426 } (\bibinfo{year}{1982}).

\bibitem[{\citenamefont{Repp et~al.}(2004)\citenamefont{Repp, Meyer, and
  Rieder}}]{Repp:04.1}
\bibinfo{author}{\bibfnamefont{J.}~\bibnamefont{Repp}},
  \bibinfo{author}{\bibfnamefont{G.}~\bibnamefont{Meyer}}, \bibnamefont{and}
  \bibinfo{author}{\bibfnamefont{K.-H.} \bibnamefont{Rieder}},
  \bibinfo{journal}{Phys. Rev. Lett.} \textbf{\bibinfo{volume}{92}},
  \bibinfo{pages}{036803} (\bibinfo{year}{2004}).

\bibitem[{\citenamefont{H\"{o}vel et~al.}(2001)\citenamefont{H\"{o}vel, Grimm,
  and Reihl}}]{Reihl:01}
\bibinfo{author}{\bibfnamefont{H.}~\bibnamefont{H\"{o}vel}},
  \bibinfo{author}{\bibfnamefont{B.}~\bibnamefont{Grimm}}, \bibnamefont{and}
  \bibinfo{author}{\bibfnamefont{B.}~\bibnamefont{Reihl}},
  \bibinfo{journal}{Surface Science} \textbf{\bibinfo{volume}{477}},
  \bibinfo{pages}{43} (\bibinfo{year}{2001}).

\bibitem[{\citenamefont{Yokoyama et~al.}(2007)\citenamefont{Yokoyama,
  Takahashi, Shinozaki, and Okamoto}}]{Yokoyama:07.5}
\bibinfo{author}{\bibfnamefont{T.}~\bibnamefont{Yokoyama}},
  \bibinfo{author}{\bibfnamefont{T.}~\bibnamefont{Takahashi}},
  \bibinfo{author}{\bibfnamefont{K.}~\bibnamefont{Shinozaki}},
  \bibnamefont{and} \bibinfo{author}{\bibfnamefont{M.}~\bibnamefont{Okamoto}},
  \bibinfo{journal}{Phys. Rev. Lett.} \textbf{\bibinfo{volume}{98}},
  \bibinfo{pages}{206102} (\bibinfo{year}{2007}).

\bibitem[{\citenamefont{Fernandez-Torrente
  et~al.}(2007)\citenamefont{Fernandez-Torrente, Monturet, Franke, Fraxedas,
  Lorente, and Pascual}}]{Pascual:07.10}
\bibinfo{author}{\bibfnamefont{I.}~\bibnamefont{Fernandez-Torrente}},
  \bibinfo{author}{\bibfnamefont{S.}~\bibnamefont{Monturet}},
  \bibinfo{author}{\bibfnamefont{K.~J.} \bibnamefont{Franke}},
  \bibinfo{author}{\bibfnamefont{J.}~\bibnamefont{Fraxedas}},
  \bibinfo{author}{\bibfnamefont{N.}~\bibnamefont{Lorente}}, \bibnamefont{and}
  \bibinfo{author}{\bibfnamefont{J.~I.} \bibnamefont{Pascual}},
  \bibinfo{journal}{Phys. Rev. Lett.} \textbf{\bibinfo{volume}{99}},
  \bibinfo{pages}{176103} (\bibinfo{year}{2007}).

\bibitem[{\citenamefont{Mehlhorn et~al.}(2010)\citenamefont{Mehlhorn,
  {Simic-Milosevic}, Jaksch, Scheier, and Morgenstern}}]{morgenstern:10.7}
\bibinfo{author}{\bibfnamefont{M.}~\bibnamefont{Mehlhorn}},
  \bibinfo{author}{\bibfnamefont{V.}~\bibnamefont{{Simic-Milosevic}}},
  \bibinfo{author}{\bibfnamefont{S.}~\bibnamefont{Jaksch}},
  \bibinfo{author}{\bibfnamefont{P.}~\bibnamefont{Scheier}}, \bibnamefont{and}
  \bibinfo{author}{\bibfnamefont{K.}~\bibnamefont{Morgenstern}},
  \bibinfo{journal}{Surface Science} \textbf{\bibinfo{volume}{604}},
  \bibinfo{pages}{1698} (\bibinfo{year}{2010}).

\bibitem[{\citenamefont{Wray et~al.}(2011)\citenamefont{Wray, Xu, Xia, Hsieh,
  Fedorov, Hor, Cava, Bansil, Lin, and Hasan}}]{Hasan:11.1}
\bibinfo{author}{\bibfnamefont{L.~A.} \bibnamefont{Wray}},
  \bibinfo{author}{\bibfnamefont{S.}~\bibnamefont{Xu}},
  \bibinfo{author}{\bibfnamefont{Y.}~\bibnamefont{Xia}},
  \bibinfo{author}{\bibfnamefont{D.}~\bibnamefont{Hsieh}},
  \bibinfo{author}{\bibfnamefont{A.~V.} \bibnamefont{Fedorov}},
  \bibinfo{author}{\bibfnamefont{Y.~S.} \bibnamefont{Hor}},
  \bibinfo{author}{\bibfnamefont{R.~J.} \bibnamefont{Cava}},
  \bibinfo{author}{\bibfnamefont{A.}~\bibnamefont{Bansil}},
  \bibinfo{author}{\bibfnamefont{H.}~\bibnamefont{Lin}}, \bibnamefont{and}
  \bibinfo{author}{\bibfnamefont{M.~Z.} \bibnamefont{Hasan}},
  \bibinfo{journal}{Nat Phys} \textbf{\bibinfo{volume}{7}}, \bibinfo{pages}{32}
  (\bibinfo{year}{2011}).

\bibitem[{\citenamefont{Zhu et~al.}(2011)\citenamefont{Zhu, Yao, Zhang, and
  Chang}}]{Chang:11.2}
\bibinfo{author}{\bibfnamefont{J.-J.} \bibnamefont{Zhu}},
  \bibinfo{author}{\bibfnamefont{D.-X.} \bibnamefont{Yao}},
  \bibinfo{author}{\bibfnamefont{S.-C.} \bibnamefont{Zhang}}, \bibnamefont{and}
  \bibinfo{author}{\bibfnamefont{K.}~\bibnamefont{Chang}},
  \bibinfo{journal}{Phys. Rev. Lett.} \textbf{\bibinfo{volume}{106}},
  \bibinfo{pages}{097201} (\bibinfo{year}{2011}).

\end{thebibliography}

\end{document}